\documentclass[12pt]{article}
\usepackage{amssymb,amsmath,epsfig}

\begin{document}
\title{\bf Thermodynamic Behavior of particular $f(R,T)$ Gravity Models}

\author{M. Sharif \thanks{msharif.math@pu.edu.pk} and M. Zubair
\thanks{mzubairkk@gmail.com}\\\\
Department of Mathematics, University of the Punjab,\\
Quaid-e-Azam Campus, Lahore-54590, Pakistan.}

\date{}

\maketitle

\begin{abstract}
We investigate the thermodynamics at the apparent horizon of the FRW
universe in $f(R,T)$ theory under non-equilibrium description. The
laws of thermodynamics have been discussed for two particular models
of $f(R,T)$ theory. The first law of thermodynamics is expressed in
the form of Clausius relation $T_hd\hat{S}_h=\delta{Q}$, where
$\delta{Q}=-d\hat{E}+Wd\mathbb{V}+T_hd_{\jmath}\hat{S}$ is the
energy flux across the horizon and $d_{\jmath}\hat{S}$ is the
entropy production term. Furthermore, the conditions to preserve the
generalized second law of thermodynamics are established with the
constraints of positive temperature and attractive gravity. We have
illustrated our results for some concrete models in this theory.
\end{abstract}

\section{Introduction}

Recent astrophysical observations indicate that expansion of the
universe is presently in an accelerated epoch. The most compelling
evidence for this is found in measurements of supernovae type Ia
(SNeIa) \cite{1} which is supported by renowned observations
\cite{4}-\cite{8}. The mysterious component of energy named as
\emph{dark energy} (DE) is often introduced to explain this behavior
of the universe. However, the mechanism responsible for the
accelerated expansion is still under debate.

Two approaches have been used to illustrate the issue of current
cosmic acceleration. Introducing an ``exotic cosmic fluid" in the
framework of Einstein gravity \cite{10}-\cite{12} is one direction
to deal such issue, however this approach did not fully explain
current empirical data. The other way is to discuss the modified
theories of gravity such as $f(R)$ \cite{16,17}, $f(\mathcal{T})$
\cite{18}, where "$\mathcal{T}$" is the torsion scalar in
teleparallel and $f(R,T)$, where $R$ and $T$ being the Ricci sclar
and the trace of energy-momentum tensor \cite{21}-\cite{21a} etc.
The $f(R,T)$ theory modifies Einstein Lagrangian through the
coupling of matter and geometry. In fact, this modified gravity
generalizes the $f(R)$ theory and necessitates an arbitrary function
of $R$ and $T$. Recently, Bamba et al. \cite{22} presented a
comprehensive review of the problem of DE and modified theories.

Black hole thermodynamics suggests that there is a fundamental
connection between gravitation and thermodynamics \cite{23}. Hawking
radiations \cite{24} together with; proportionality relation between
temperature and surface gravity, also connection between horizon
entropy and area of a black hole \cite{25} further support this
idea. Jacobson \cite{26} was the first to deduce the Einstein field
equations from the Clausius relation $T_hd\hat{S}_h={\delta}\hat{Q}$
together with entropy proportional to the horizon area. In case of a
general spherically symmetric spacetime, it was shown that the field
equations can be constituted as the first law of thermodynamics
(FLT) \cite{28}.

The relation between the FRW equations and the FLT  was shown in
\cite{30} for $T_h=1/2{\pi}\tilde{r}_A,~ S_h=\pi{\tilde{r}^2_A}/G$.
The field equations for FRW background were also formulated in
\emph{Gauss-Bonnet} and \emph{Lovelock} theories by employing the
corresponding entropy relation of static spherically symmetric black
holes. Eling et al. \cite{31} shown that we cannot find the correct
field equations simply by using the Clausius relation in nonlinear
theories of gravity. They remarked that the non-equilibrium
description of thermodynamics is needed, whereby the Clausius
relation is modified to $T_hdS_h=\delta{Q}+d_{\jmath}S$, where
$d_{\jmath}S$ is the entropy production term. In
ref.\cite{32}-\cite{40} it was shown that the FRW field equations in
general relativity (GR) and modified theories can be rewritten as
$dE=T_hdS_h+WdV$ (unified FLT on the trapping horizon suggested by
Hayward \cite{34}) with work term $W=\frac{1}{2}(\rho-p)$.

Wu et al. \cite{41} developed a generalized procedure to construct
FLT and the generalized second law of thermodynamics (GSLT) at the
apparent horizon of Friedmann universe. The validity conditions of
GSLT were studied in modified theories of gravity. Gong and Wang
\cite{43} showed that equilibrium thermodynamics is achievable for
extended theories of gravity and entropy correction terms can be
confined to mass-like functions. Other alternative approaches
\cite{44}-\cite{48} have also been developed to reinterpret the
non-equilibrium correction. In \cite{53}, we have explored the GSLT
in $f(R,T)$ theory and found necessary conditions for its validity.
It has been shown that the equilibrium description isn't feasible by
redefining the \emph{dark energy components} in $f(R,T)$ theory.

In present work, the thermodynamics laws are examined for two
particular models of $f(R,T)$ theory. We show that the FRW equations
can be rewritten in a from of the FLT
$T_hd\hat{S}_h+T_hd_{\jmath}\hat{S}_h=-d\hat{E}+W_{tot}dV$. We
formulate the GSLT and explore the conditions to validate this law.
The paper is arranged in following format: In section \textbf{2}, we
present a brief introduction of $f(R,T)$ theory. Section \textbf{3}
is devoted to discuss the FLT and GSLT corresponding to the
Friedmann equations of particular $f(R,T)$ models. Finally,
concluding remarks are given in section \textbf{4}.

\section{$f(R,T)$ Gravity: An Overview}

The $f(R,T)$ modified gravity is described by the action \cite{21}
\begin{equation}\label{1}
\mathcal{I}=\int{dx^4\sqrt{-g}\left[\frac{f(R,T)}{2\kappa}+\mathcal{L}_{m}\right]},
\end{equation}
where $\kappa=8\pi{G}$, $\mathcal{L}_{m}$ defines the matter
substances of the universe. The matter energy-momentum tensor
$T_{\alpha\beta}^{(m)}$ is defined as \cite{55}
\begin{equation}\label{2}
T^{(m)}_{\alpha\beta}=-\frac{2}{\sqrt{-g}}\frac{\delta(\sqrt{-g}
{\mathcal{\mathcal{L}}_{m}})}{\delta{g^{\alpha\beta}}}.
\end{equation}
The field equations can be found by varying the action of $f(R,T)$
gravity with respect to the metric tensor
\begin{eqnarray}\label{3}
&&R_{\alpha\beta}f_{R}(R,T)-\frac{1}{2}g_{\alpha\beta}f(R,T)+(g_{\alpha\beta}
{\Box}-{\nabla}_{\alpha}{\nabla}_{\beta})f_{R}(R,T)\nonumber\\&=&8{\pi}G
T^{(m)}_{\alpha\beta}-f_{T}(R,T)T^{(m)}_{\alpha\beta}-f_{T}(R,T)\Theta_{\alpha\beta},
\end{eqnarray}
where $f_{R}$ and $f_{T}$ represent derivatives of $f(R,T)$ with
respect to $R$ and $T$, respectively. The field equations depend on
the source term $\Theta_{\mu\nu}$, hence every selection of
${\mathcal{L}}_{m}$ generates a particular set of field equations.

We consider the perfect fluid as matter source with matter
Lagrangian $\mathcal{L}_{m}=p_m$ so that $\Theta_{\alpha\beta}$ is
given by
\begin{equation}\label{5}
\Theta_{\alpha\beta}=-2T_{\alpha\beta}^{(m)}+p_{m}g_{\alpha\beta}.
\end{equation}
Substituting this value in Eq.(\ref{3}), it follows that
\begin{eqnarray}\label{6}
R_{\alpha\beta}f_{R}-\frac{1}{2}g_{\alpha\beta}f+(g_{\alpha\beta}
{\Box}-{\nabla}_{\alpha}{\nabla}_{\beta})f_{R}=8{\pi}G
T_{\alpha\beta}^{(m)}+T_{\alpha\beta}^{(m)}f_{T}-p_{m}
g_{\alpha\beta}f_{T}.
\end{eqnarray}
The spatially homogeneous and isotropic, ($n+1$)-dimensional FRW
universe is defined as
\begin{equation}\label{9}
ds^{2}=h_{\alpha\beta}dx^{\alpha}dx^{\beta}+\tilde{r}^{2}d{\hat{\Omega}}_{n-1}^2,
\end{equation}
where $h_{\alpha\beta}=diag(-1,a^2/(1-kr^2))$ is the 2-dimensional
metric, $a(t)$ is the scale factor and $k$ is the cosmic curvature;
$\tilde{r}=a(t)r,~x^{0}=t,~x^{1}=r$ and $d{\hat{\Omega}}_{n-1}^2$ is
the metric of a ($n-1$)-dimensional sphere. For $n=3$, we have
($3+1$)-dimensional FRW metric in Einstein gravity, while one can
have $n\geq4$ in other gravity theories.

\section{Thermodynamics in $f(R,T)$ gravity}

Now, we discuss the laws of thermodynamics for two particular
choices of $f(R,T)$ gravity \cite{21}.

\subsection{$f(R,T)=f_{1}(R)+f_{2}(T)$}

Let us consider the following $f(R,T)$ model
\begin{equation}\label{13}
f(R,T)=f_{1}(R)+f_{2}(T),
\end{equation}
where $f_{1}$ and $f_{2}$ are arbitrary functions of $R$ and $T$,
respectively. The corresponding field equations are
\begin{eqnarray}\label{14}
&&R_{\alpha\beta}f_{1R}(R)-\frac{1}{2}g_{\alpha\beta}f_{1}(R)+(g_{\alpha\beta}
{\Box}-{\nabla}_{\alpha}{\nabla}_{\beta})f_{1R}(R)\nonumber\\&=&8{\pi}G
T_{\alpha\beta}^{(m)}+T_{\alpha\beta}^{(m)}f_{2T}(T)+\frac{1}{2}
g_{\alpha\beta}f_{2}(T),
\end{eqnarray}
where $f_{1R}(R)=df_1/dR$ and $df_{2}/dT$. The choice of
$f_{2}(T)=0$ implies the field equation of $f(R)$ gravity. In FRW
background, the field equations will become
\begin{eqnarray}\label{15}
\left(H^2+\frac{k}{a^2}\right)&=&\frac{16{\pi}G_{Eff}}{n(n-1)}
(\rho_m+\rho_{dc}),\\\label{16}\left(\dot{H}-\frac{k}{a^2}\right)
&=&-\frac{8{\pi}G_{Eff}}{(n-1)}({\rho}_m+\rho_{dc}+p_{dc}),
\end{eqnarray}
where $G_{Eff}=\frac{1}{f_{1R}}\left(G+\frac{f_{2T}}{8\pi}\right)$,
and
\begin{eqnarray}\label{17}
\rho_{dc}&=&\frac{1}{8{\pi}G\mathcal{D}}\left[\frac{1}{2}(Rf_{1R}
-f_1-f_2)-nH\dot{R}f_{1RR}\right],\\\nonumber{p}_{dc}&=&\frac{1}
{8{\pi}G\mathcal{D}}\left[-\frac{1}{2}(Rf_{1R}-f_1-f_2)+(n-1)H\dot{R}
f_{1RR}+\ddot{R}f_{1RR}\right.\\\label{18}&+&\left.\dot{R}^2f_{1RRR}\right]
\end{eqnarray}
and $\mathcal{D}=\left(1+\frac{f_{2T}(R,T)}{8{\pi}G}\right)$.
Substituting Eqs.(\ref{17}) and (\ref{18}) in conservation equation
\cite{53}, we obtain
\begin{equation}\label{19}
q_{t}=\frac{n(n-1)}{16{\pi}G}(H^2+\frac{k}{a^2})\partial_t
\left(\frac{f_{1R}}{\mathcal{D}}\right).
\end{equation}
Clearly, this reduces to the energy transfer relation for $f(R)$
theory if $f(R,T)=f_1(R)$ \cite{47,48}. If the effective
gravitational coupling is constant, we obtain $q_{t}=0$.

\subsubsection{First Law of Thermodynamics}

Now, the FLT is constructed for the above $f(R,T)$ model. The
condition,
$h^{\upsilon\lambda}\partial_{\upsilon}\tilde{r}\partial_{\lambda}\tilde{r}=0$,
gives the radius $\tilde{r}_A$ of the apparent horizon
\begin{equation*}
\tilde{r}_A=\left(H^2+\frac{k}{a^2}\right)^{-1/2}.
\end{equation*}
The associated temperature is $T_h=\frac{|\kappa_{sg}|}{2\pi}$,
where $\kappa_{sg}=\frac{1}{2\sqrt{-h}}\partial_{\mu}(\sqrt{-h}h^
{\mu\nu}\partial_{\nu}\tilde{r}_A)=-\frac{1}{\tilde{r}_A}
(1-\frac{1}{2H}\frac{d[ln\tilde{r}_A]}{dt})$ is the surface gravity
\cite{30}. The temperature $T_h=\frac{1}{2{\pi}\tilde{r}_A}(1-\eta)$
is positive for $\eta=\frac{1}{2H}\frac{d[ln\tilde{r}_A]}{dt}<1$.
Applying the definition of $\tilde{r}_A$, the condition to keep
$T_h$ positive is expressed as
\begin{equation}\label{22}
(\dot{H}-\frac{k}{a^2})>-2(H^2+\frac{k}{a^2}).
\end{equation}

In GR, the horizon entropy is defined as $S_h=A/4G$
\cite{23}-\cite{25}, where
$A=n\hat{\Omega}_n\tilde{r}^{n-1}_A=n\pi^{n/2}[\Gamma(n/2+1)]^{-1}\tilde{r}^{n-1}_A$
represents area of the apparent horizon. Wald \cite{56} proposed
that in modified gravitational theories, the horizon entropy is
associated with a Noether charge entropy. Brustein et al. \cite{57}
showed that Wald entropy is equivalent to $S_h=A/4G_{Eff}$ where
$G_{Eff}$ being the effective gravitational coupling. We can define
the Wald entropy in $f(R,T)$ theory as \cite{53}
\begin{equation}\label{23}
\hat{S}_h=\frac{A}{4G_{Eff}},
\end{equation}
where $G_{Eff}=G\mathcal{D}(R,T)/f_{1R}$ for the first $f(R,T)$
model. Following ref.\cite{53}, one can obtain the FLT in the
following form
\begin{equation*}
T_hd\hat{S}_h=\delta{Q},
\end{equation*}
where the energy flux $\delta{Q}$ is
\begin{eqnarray}\label{31}
\delta{Q}&=&-d\hat{E}+\frac{n}{2}\tilde{r}_{A}^{n-1}(\rho_{t}
-p_{t})d\tilde{r}_A+\frac{n\hat{\Omega}_n(n-1)\tilde{r}^{n-2}_A}{16{\pi}G}
d\left(\frac{f_{1R}}{\mathcal{F}}\right)\nonumber\\
&=&-d\hat{E}+W_{t}d\mathbb{V}+\mathbb{V}q_{t}dt+T_hS_hd\left(\frac{f_{1R}}{\mathcal{D}}\right).
\end{eqnarray}
$W_{t}=-\frac{1}{2}T^{(t)\mu\nu}h_{\mu\nu}=\frac{1}{2}
(\rho_{t}-p_{t})$ is the total work density \cite{34}. Thus, the FLT
can be expressed as
\begin{equation}\label{32}
T_hd\hat{S}_h+T_hd_{\jmath}\hat{S}_h=-d\hat{E}+W_{tot}d\mathbb{V},
\end{equation}
where
\begin{equation*}
d_{\jmath}\hat{S}_h=-\frac{n\hat{\Omega}_n(H^2+\frac{k}{a^2})^\frac{1-n}{2}((n+1)H^2+\dot{H}+
n\frac{k}{a^2})d(f_{1R}/\mathcal{D})}{4G(2H^2+\dot{H}+\frac{k}{a^2})}
\end{equation*}
is the entropy production term developed for this model. This
characterizes non-equilibrium treatment of the thermodynamics. The
FLT for a flat FRW universe in $f(R)$ theory \cite{47,48} can be
retrieved from this result. For $f(R,T)=R$, the term
$d_{\jmath}\hat{S}_h$ vanishes, which leads to the FLT in Einstein
gravity.

\subsubsection{Generalized Second Law of Thermodynamics}

Now we investigate the validity of GSLT in $f(R,T)$ theory for this
model. The FLT determines the horizon entropy given by
Eq.(\ref{32}). The composition of the entire matter and energy
fluids within the horizon is given by Gibb's equation \cite{59}
\begin{equation}\label{35}
T_{t}d\hat{S}_{t}=dE_{t}+p_{t}d\mathbb{V},
\end{equation}
where $T_{t}$ and $\hat{S}_{t}$ represent the temperature and
entropy of all contents within the horizon, respectively. There is a
relation of temperature within the horizon to $T_h$ \cite{41}
\emph{i.e.}, $T_{t}=bT_h$, here $0<b<1$ to assure that $0<T_t<T_h$.
Consider $\hat{S}$ being the sum of matter entropy within the
horizon, horizon entropy and the non-equilibrium entropy production
term.

The GSLT states that the time derivative of total entropy is not
decreasing with time \emph{i.e.,}
\begin{equation}\label{36}
T_h\dot{\hat{S}}=T_h(\dot{\hat{S}}_h+d_\jmath\dot{\hat{S}}_h+\dot{\hat{S}}_{t})\geqslant0,
\end{equation}
where $d_\jmath\dot{\hat{S}}_h=\partial_t(d_\jmath\hat{S}_h)$.
Inserting Eqs.(\ref{32}) and (\ref{35}) in the above inequality, we
obtain
\begin{eqnarray}\nonumber
&&\frac{n(n-1)\hat{\Omega}_n}{16{\pi}HG}\left[2H\dot{\tilde{r}}_{A}
\left[(b-1)+\dot{\tilde{r}}_{A}\tilde{r}_{A}^{n-4}(2-b)\right]\left(\frac{f_{1R}}{\mathcal{D}}\right)
+(1-b)H\tilde{r}_{A}\right.\\\label{37}&\times&\left. \partial_t
\left(\frac{f_{1R}}{\mathcal{D}}\right)\right]\geqslant0, \quad
where \quad \quad
\dot{\tilde{r}}_{A}=-\tilde{r}^{-3}_{A}H(\dot{H}-\frac{\kappa}{a^2}).
\end{eqnarray}
We can impose the constraint $\mathcal{D}/f_{1R}>0$ so that
$G_{Eff}$ is positive. Applying the positive temperature condition
$(\dot{H}-\frac{k}{a^2})>-2(H^2+\frac{k}{a^2})$ with the temperature
parameter $b<1$, this gives
\begin{eqnarray}\label{39}
&&\frac{n(n-1)\hat{\Omega}_n(H^2+\frac{k}{a^2})^{-\frac{n}{2}+1}}{16{\pi}G\mathcal{D}}
\left[4Hf_{1R}+(1-b){\mathcal{D}}\partial_t\left(\frac{f_{1R}}
{\mathcal{D}}\right)\right]\geqslant0.
\end{eqnarray}

Thus the GSLT can be satisfied, provided that
$\partial_t(f_{1R}/\mathcal{D})>0$. If
$\partial_t(f_{1R}/\mathcal{D})<0$, then the GSLT is protected only
if
$\mid\frac{\partial_t(f_{1R}/\mathcal{D})}{f_{1R}/\mathcal{D}}\mid
\leqslant\frac{4H}{1-b}$. If the gravitational coupling constant is
indeed a constant \emph{i.e.,} $\partial_t(f_{1R}/\mathcal{D})=0$,
then the GSLT always holds. The condition to preserve the GSLT in
$f(R)$ theory can be reproduced if $f(R,T)=f_1(R)$. For $k=0$ and
$f_2(T)=0$, one get the inequality already constructed by Wu et al.
\cite{41} in non-linear gravity. In the thermal-equilibrium limit
$b\sim1$, the constraint to protect the GSLT is
\begin{eqnarray}\label{39a}
\frac{n(n-1)\hat{\Omega}_n(H^2+\frac{k}{a^2})^{-(\frac{n}{2}+1)}}
{16{\pi}G\mathcal{D}}\left[H(\dot{H}-\frac{k}{a^2})^2f_{1R}\right]
\geqslant0.
\end{eqnarray}
The relation (\ref{39a}) depends on the choice of $f(R,T)$, for
instance $f_1(R)=R$ with $f_2(T)=0$ results in
\begin{eqnarray}\nonumber
\frac{n(n-1)\hat{\Omega}_n(H^2+\frac{k}{a^2})^{-(\frac{n}{2}+1)}}
{16{\pi}G}\left[H(\dot{H}-\frac{k}{a^2})^2\right] \geqslant0.
\end{eqnarray}
which is the condition for validity of GSLT in Einstein gravity.
Here, we discuss the validity of GSLT for some particular forms of
$f(R,T)$ gravity namely,\\
(i) $f_1(R)=f(R),\quad f_2(T)=\lambda{T}$,\\
(ii) $f_1(R)=R,\quad f_2(T)=2f(T)$.\\
In the first case, we consider the $f(R,T)$ model corresponding to
the power law solution $a(t)=a_0t^m$ \cite{21a}
\begin{equation}\label{39b}
f(R,T)=\alpha_{k\omega}(-R)^k+\lambda{T},
\end{equation}
where
\begin{equation}\nonumber
\alpha_{k\omega}=\frac{2^{3-2k}3^{k-1}kA(k(4k-3(1+\omega))^{1-k}
(1+\omega)^{2k-2}}{k^2(6\omega+8)-k(9\omega+13)+3(\omega+1)}.
\end{equation}
For this model, the Hubble and deceleration parameters are
$H=\frac{2k}{3(1+\omega)}$ and $q=-1+\frac{3(1+\omega)}{2k}$. The
validity of the GSLT in ($3+1$)-dimensional flat FRW universe for
the model (\ref{39b}) requires the condition
\begin{equation}\label{39c}
T_h\dot{\hat{S}}=\frac{9(1+\omega)^2\alpha_{k\omega}}{8k^2\tilde{G}}
\left(\frac{4k[4k-3(1+\omega)]}{3(1+\omega)^2t^2}\right)^{k-1}\geqslant0,
\end{equation}
where $\tilde{G}=G+\frac{\lambda}{8\pi}$. Now we present some
constraints for particular values of $k={-2,-1,1,2}$.
\begin{itemize}
\item {For $k=1$ this solution represents $\Lambda$CDM model and
constraint on GSLT is given by\\
$T_h\dot{\hat{S}}=\frac{9(1+\omega)^2A}{8\tilde{G}}\geqslant0$,
which is true if $A>0$ with ${\omega}\leqslant3$.}
\item {For $k=2$, the GSLT is valid if
$T_h\dot{\hat{S}}=\frac{A(5-3\omega)^2}{2\tilde{G}t^2}\geqslant0$,
which requires $A<0$.}
\item {For $k=-1,-2$, we find $T_h\dot{\hat{S}}\geqslant0$ if $A>0$ with $w\geqslant0$.
This choice would favor the expanding universe since $q<-1$.}
\end{itemize}
The higher powers of curvature can be made available for larger
values of $k$ and one can examine the validity of GSLT. If one
consider the dust case $\omega=0$, then the possible role of
$\lambda$ and $k$ can be seen from graphical description as shown in
Figure \textbf{1}.
\begin{figure}
\centering \epsfig{file=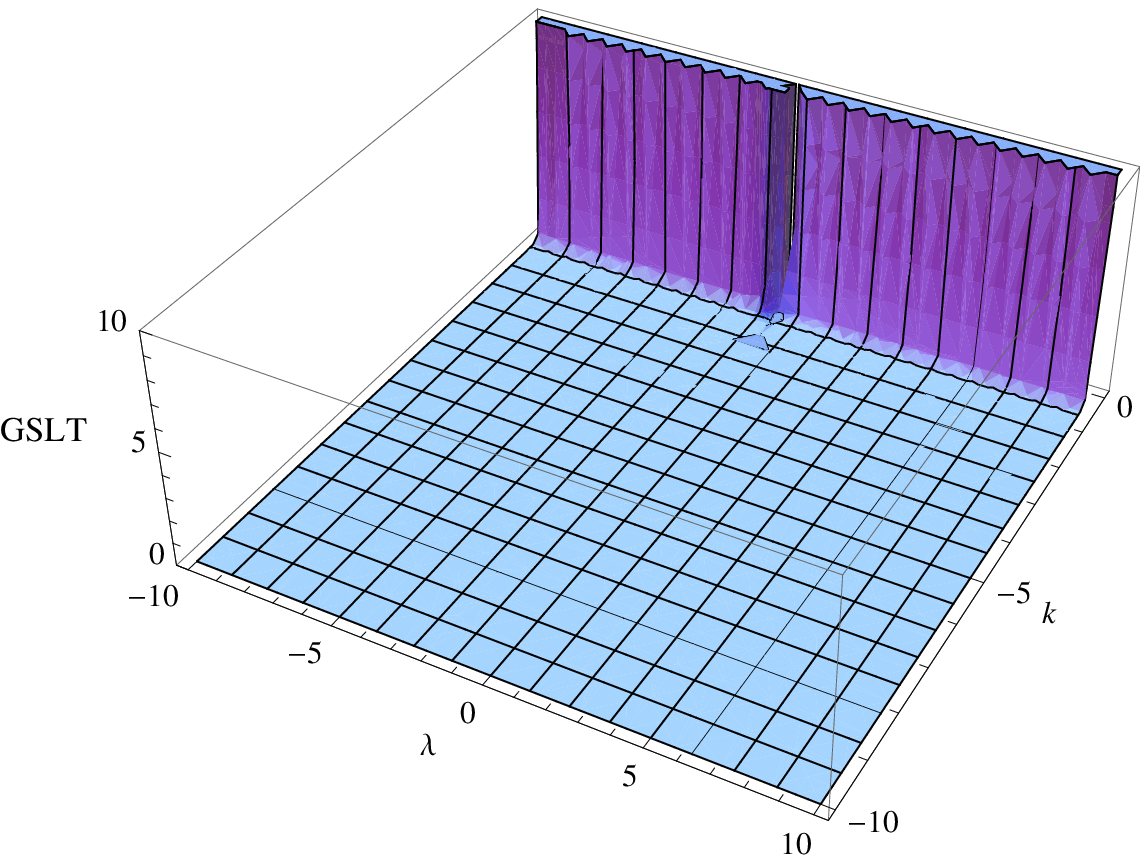, width=.495\linewidth,
height=2.2in} \epsfig{file=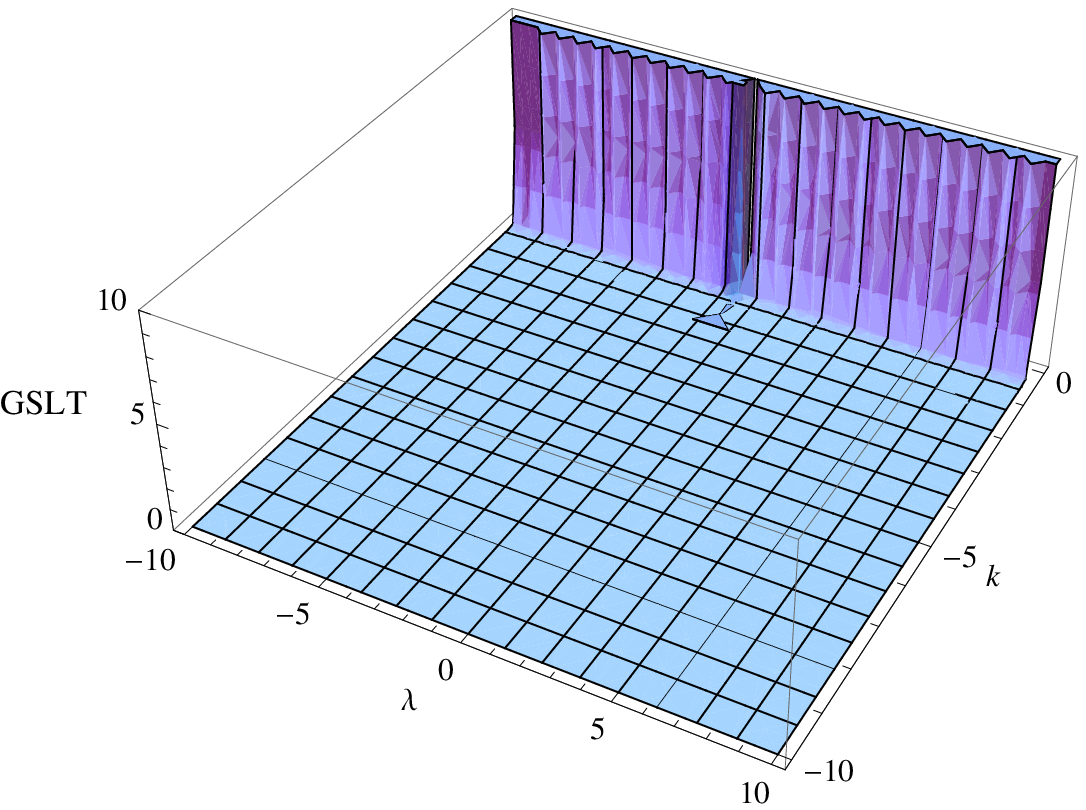, width=.495\linewidth,
height=2.2in} \caption{Evolution of GSLT for different values of
parameters $k$ and $\lambda$ (a) for present epoch $z=0$ and (b) for
$z=-0.9$}
\end{figure}

In the second case, the GSLT for the $f(R,T)=R+2f(T)$ model demands
the following inequality to be fulfilled
\begin{equation}\nonumber
T_h\dot{\hat{S}}=\frac{\dot{H}^2}{2H^4\hat{G}}\geqslant0, \quad
where \quad \hat{G}=G+2f(T)/8{\pi}.
\end{equation}
Here we consider the power law solution of the form
$f(T)=a_1T+a_2T^k$, where $a_1$ and $a_2$ are parameters. Following
\cite{21a} for the dust case, we set $a_1=1$ and
$a_2=\frac{2^{3-2k}3^{k-1}k^{3-2k}}{4+2k}$. Thus the above
inequality takes the form
\begin{equation}\nonumber
T_h\dot{\hat{S}}=\frac{9\pi(4+2k)}{k^2[(4+2k)(8\pi{G}+1)+2^{3-2k}3^{k-1}k^{4-2k}T^{k-1}]}\geqslant0.
\end{equation}
This shows that the GSLT holds for the $f(T)$ power law model and
its validity is shown in Figure \textbf{2}.
\begin{figure}
\centering \epsfig{file=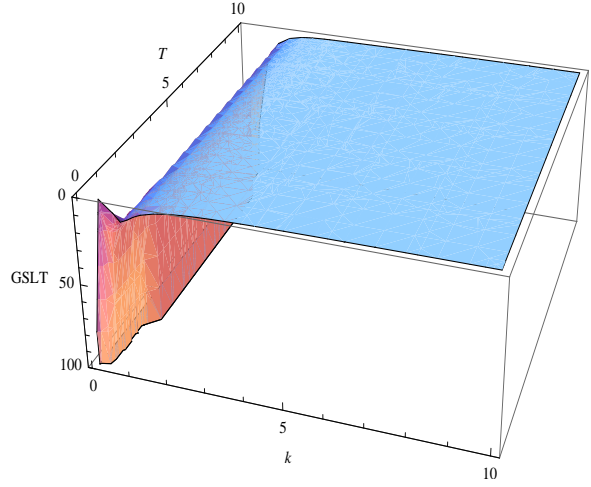, width=.6\linewidth,
height=2.5in}\caption{ Evolution of GSLT for case (ii) versus $T$
and $k$.}
\end{figure}

\subsection{$f(R,T)=f_{1}(R)+f_2(R)f_{3}(T)$}

A more general $f(R,T)$ gravity model is of the form \cite{21}
\begin{equation}\label{40}
f(R,T)=f_{1}(R)+f_2(R)f_{3}(T),
\end{equation}
where $f_{i}~(i=1,2)$ are functions of $R$ and $f_3$ is function of
$T$. For a dust matter source, the field equation is obtained as
\begin{eqnarray}\label{41}
&&R_{\alpha\beta}[f_{1R}+f_{2R}f_{3}]-\frac{1}{2}g_{\alpha\beta}
f_{1}+(g_{\alpha\beta}{\Box}-{\nabla}_{\alpha}{\nabla}_{\beta})[f_{1R}
+f_{2R}f_{3}]\nonumber\\&=&8{\pi}GT_{\alpha\beta}^{(m)}+
T_{\alpha\beta}^{(m)}f_{2}f_{3T}+\frac{1}{2}
g_{\alpha\beta}f_{2}f_{3}.
\end{eqnarray}
An equivalent Einstein field equation can be obtained with
$G_{Eff}=\frac{1}{f_{1R}+f_{2R}f_3}\left(G+\frac{f_2f_{3T}}{8\pi}\right)$,
whereas
\begin{eqnarray}\nonumber
&&\hat{T}_{\alpha\beta}^{(dc)}=\frac{1}{f_{1R}+f_{2R}f_3}\left[\frac{1}{2}g_{\alpha\beta}
(f_{1}+f_{2}f_3)-R(f_{1R}+f_{2R}f_3))+({\nabla}_{\alpha}{\nabla}_{\beta}
\right.\\\label{43}&-&\left.g_{\alpha\beta}{\Box})(f_{1R}+f_{2R}f_3)\right].
\end{eqnarray}
In the following discussion, we define
$\mathcal{J}(R,T)=f_{1R}(R)+f_{2R}(R)f_{3}(T)$.

For this $f(R,T)$ model, the field equation are identical to
Eqs.(\ref{15})-(\ref{16}), whereas
\begin{eqnarray}\label{46}
\hat{\rho}_{dc}&=&\frac{1}{8{\pi}G\mathcal{B}}\left[\frac{1}{2}(R\mathcal{J}
-f_1-f_2f_3)-nH(\dot{R}\mathcal{J}_R+\dot{T}\mathcal{J}_T)\right],\\\nonumber{\hat{p}}_{dc}&=&\frac{1}
{8{\pi}G\mathcal{B}}\left[-\frac{1}{2}(R\mathcal{J}-f_1-f_2f_3)+(n-1)H(\dot{R}\mathcal{J}_R
+\dot{T}\mathcal{J}_T)+\ddot{R}\mathcal{J}_{R}\right.\\\label{47}&+&\left.
\dot{R}^2\mathcal{J}_{RR}+2\dot{R}\dot{T}\mathcal{J}_{RT}+\ddot{T}\mathcal{J}_{T}
+\dot{T}^2\mathcal{J}_{TT}\right]
\end{eqnarray}
and
$\mathcal{B}(R,T)=\left(1+\frac{f_2(R)f_{3T}(T)}{8{\pi}G}\right)$
which includes contributions from both matter and geometry. The
total energy exchange term for this model is given by
\begin{equation}\label{48}
q_{t}=\frac{n(n-1)}{16{\pi}G}(H^2+\frac{k}{a^2})\partial_t\left(
\frac{\mathcal{J}}{\mathcal{B}}\right).
\end{equation}
Now we analyze the validity of the FLT and GSLT for the above model.

\subsubsection{First Law of Thermodynamics}

The Wald entropy $\hat{S}_h=A/4G_{Eff}$ for function (\ref{40})
becomes
\begin{equation}\label{50}
\hat{S}_h=\frac{n\hat{\Omega}_n\tilde{r}_{A}^{n-1}\mathcal{J}}{4G\mathcal{B}}.
\end{equation}
In this case, FLT involves the energy flux $\delta{Q}$ and entropy
production terms of the form \cite{53}
\begin{eqnarray}\label{55}
\delta{Q}&=&-d\hat{E}+\frac{n}{2}\Omega_n\tilde{r}_{A}^{n-1}(\rho_{tot}
-p_{tot})d\tilde{r}_A+\frac{n(n-1)}{16{\pi}G}
\Omega_n\tilde{r}^{n-2}_Ad\left(\frac{\mathcal{J}}{\mathcal{B}}\right)\nonumber\\
&=&-d\hat{E}+W_{tot}d\mathbb{V}+\mathbb{V}q_{tot}dt+T_h\hat{S}_hd\left(\frac{\mathcal{J}}{\mathcal{B}}\right).
\\\nonumber
d_{\jmath}\hat{S}_h&=&-\frac{1}{T_h}\mathbb{V}q_{tot}dt-S_hd\left(\frac{\mathcal{J}}
{\mathcal{B}}\right)\\\label{55a}
&=&-\frac{n\Omega_n(H^2+\frac{k}{a^2})^\frac{n-1}{2}((n+1)H^2+\dot{H}+
n\frac{k}{a^2})d(\mathcal{J}/\mathcal{B})}{4G(2H^2+\dot{H}+\frac{k}{a^2})}.
\end{eqnarray}
The $f(R,T)$ gravity model, $f(R,T)=f_1(R)+f_2(R)f_3(T)$ involves
the explicit non-minimal gravitational coupling between matter and
curvature. Results obtained using this theory would be different
from other models such as $f(R)$ theory. The coupling of matter and
geometry reveals that the matter energy-momentum tensor is no longer
conserved and there is an energy transfer between the two
components. Due to this interaction, the energy exchange term
$q_{t}$ is found to be non-zero, so the entropy production term
would be an additional term in this modified gravity. Hence, the FLT
is established in more general $f(R,T)$ gravity and entropy
production is induced in a non-equilibrium treatment of
thermodynamics \cite{31,53}. In recent papers \cite{48}, Bamba et
al. shown that the entropy production term can be incorporated by
redefinition of the field equations. However, in this theory, such
treatment is not useful as shown in \cite{53}.

\subsubsection{Generalized Second Law of Thermodynamics}

To develop the GSLT for the second model, we consider the Gibbs
equation (\ref{35}). The horizon entropy is determined from FLT. The
necessary constraint for the validity of the GSLT is shown in
Eq.(\ref{36}). For the $f(R,T)$ model (\ref{40}), we obtain
\begin{eqnarray}\nonumber
&&\frac{n(n-1)\hat{\Omega}_n}{16{\pi}HG}\left[2H\dot{\tilde{r}}_{A}
\left[(b-1)+\dot{\tilde{r}}_{A}\tilde{r}_{A}^{n-4}(2-b)\right]\left(\frac{\mathcal{J}}{\mathcal{B}}\right)
+(1-b)H\tilde{r}_{A}\right.\\\label{61}&\times&\left. \partial_t
\left(\frac{\mathcal{J}}{\mathcal{B}}\right)\right]\geqslant0, \quad
where \quad \quad
\dot{\tilde{r}}_{A}=-\tilde{r}^{-3}_{A}H(\dot{H}-\frac{\kappa}{a^2}).
\end{eqnarray}

The effective gravitational coupling constant for this model is
$G_{Eff}=G\mathcal{B}/\mathcal{J}$.  We can impose the condition
$\mathcal{B}/\mathcal{J}>0$ to keep $G_{Eff}>0$. For the positive
temperature condition
$(\dot{H}-\frac{k}{a^2})>-2(H^2+\frac{k}{a^2})$ with $b<1$,
Eq.(\ref{61}) is reduced to
\begin{eqnarray}\label{62}
&&\frac{n(n-1)\hat{\Omega}_n(H^2+\frac{k}{a^2})^{-\frac{n}{2}+1}}{16{\pi}G\mathcal{B}}
\left[4H\mathcal{J}+(1-b){\mathcal{B}}\partial_t\left(\frac{\mathcal{J}}
{\mathcal{B}}\right)\right]\geqslant0.
\end{eqnarray}
This shows that the GSLT is valid only if
$\partial_t(\mathcal{J}/\mathcal{B})>0$. If the gravitational
coupling constant is indeed a constant, the GSLT is always
protected. If $\partial_t(\mathcal{J}/\mathcal{B})<0$, then the GSLT
can hold only if
$\mid\frac{\partial_t(\mathcal{J}/\mathcal{B})}{\mathcal{J}/\mathcal{B}}\mid
\leqslant\frac{4H}{1-b}$. The GSLT in $f(R)$ theory can be retrieved
for $f_3(T)=0$. If $T_t\thicksim{T}_h$, then the condition
(\ref{62}) becomes
\begin{eqnarray}\nonumber
&&\frac{n(n-1)\hat{\Omega}_n(H^2+\frac{k}{a^2})^{-(\frac{n}{2}+1)}}{16{\pi}G\mathcal{B}}\left[
H(\dot{H}-\frac{k}{a^2})^2\mathcal{J}\right]\geqslant0.
\end{eqnarray}
\begin{figure}
\centering \epsfig{file=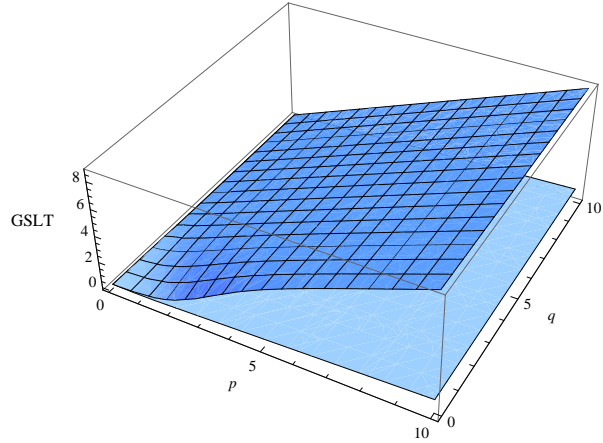, width=.6\linewidth,
height=2.3in}\caption{Evolution of GSLT for the model (36) with
$m=2$. The curve with larger slope corresponds to $z=-0.9$ while the
other represents the present value ($z=0$).}
\end{figure}
\begin{figure}
\centering \epsfig{file=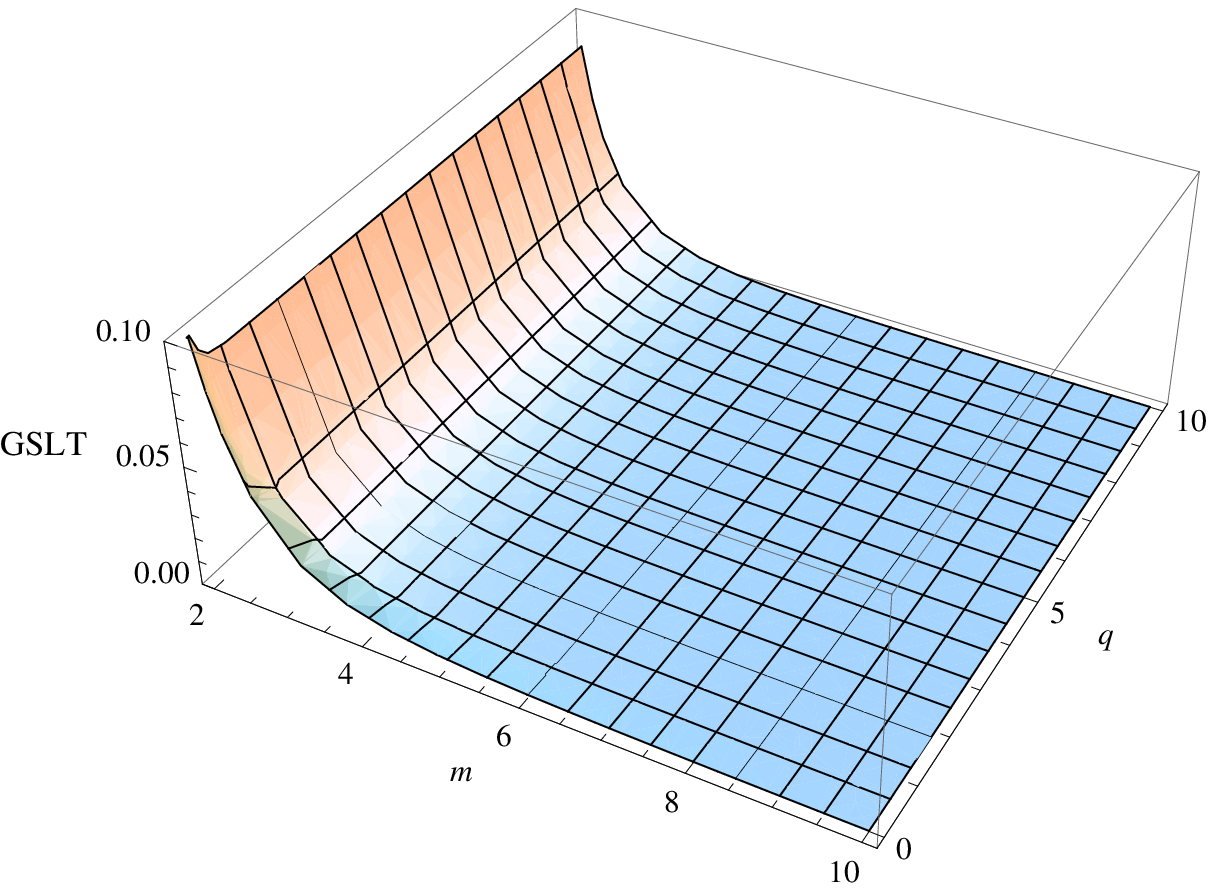, width=.495\linewidth,
height=2.2in} \epsfig{file=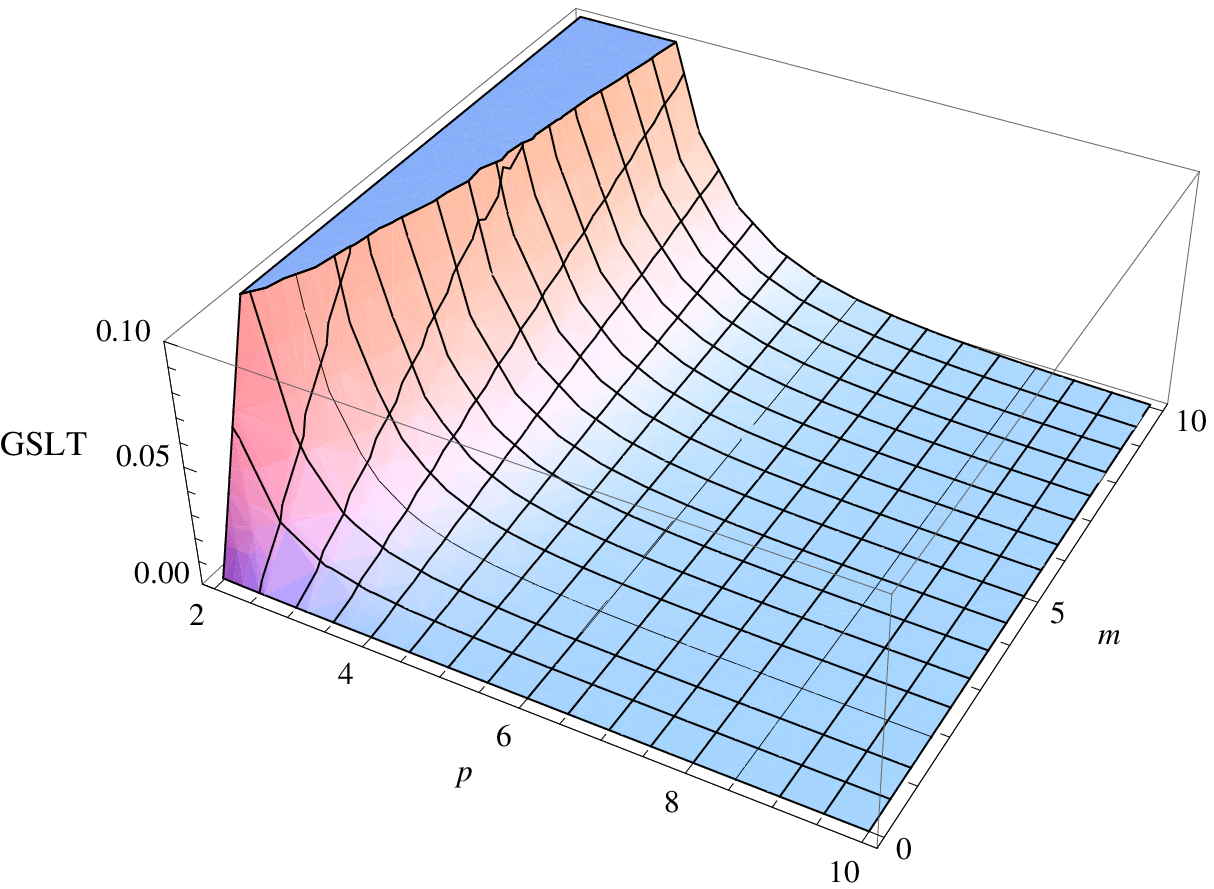, width=.495\linewidth,
height=2.2in} \caption{One can choose the specific value of one
parameter and vary the others. In left panel, we choose $p=1$ and
show the constraints on parameters $m$ and $q$. The right panel
represents the parametric values of $m$ and $p$ for fixed $q=1$.}
\end{figure}

We consider the $f(R,T)$ model (\ref{40}) with
$f_1(R)=R,~f_2(R)=R^p$ and $f_3(R)=T^q,~(p,q>0)$ so that in
4-dimensional flat FRW metric, the GSLT becomes
\begin{eqnarray}\label{63}
T_h\dot{\hat{S}}=\frac{\dot{H}^2(1+pR^{p-1}T^q)}{2H^4G\left(1+\frac{R^pT^q}{8\pi{G}}\right)}\geqslant0.
\end{eqnarray}
For the power law case $a(t)=a_0t^m$ with $\rho=\rho_0a^{-3}$, this
can be written as
\begin{eqnarray}\label{63}
T_h\dot{\hat{S}}=\frac{8\pi[1+p(6m(2m-1)t^{-2}]^{p-1}(\rho_0t^{-3m})^q}
{2m^2[8\pi{G}+(6m(2m-1)t^{-2})^{p-1}(\rho_0t^{-3m})^q]}\geqslant0.
\end{eqnarray}
We have examined the validity of relation (\ref{63}) and developed
constraints on the parameters $m$, $p$ and $q$. The results are
shown in Figures \textbf{3} and \textbf{4}.

\section{Conclusions}

In this paper, the thermodynamics properties have been discussed in
more general $f(R,T)$ theory. The non-equilibrium treatment of
thermodynamics is addressed for two particular models of $f(R,T)$
gravity. In this modified theory, accelerated expansion may produce
not just from scalar curvature part to the entire energy density of
universe, but may include a matter component as well. The
consequences of $f(R,T)$ theory may contribute to significant
results when compared to other modified gravitational theories,
applicable to various problems of contemporary interest such as
accelerated cosmic expansion, gravitational collapse, dark matter
and the detection of gravitational waves \cite{62}. The detection of
gravitational waves could be an excellent way to test general
relativity and modified theories of gravity. Corda \cite{62} has
investigated the detection of gravitational waves in $f(R)$ theory
and it would be appealing to explore this issue in $f(R,T)$ theory.

It is shown that representation of equilibrium thermodynamics is not
executable in this theory \cite{53}. Hence the non-equilibrium
treatment of thermodynamics is employed to discuss the laws of
thermodynamics. Here, we studied two particular models of $f(R,T)$
theory to show the consequences of explicit coupling of matter and
geometry. The gravitational coupling between matter and higher
derivatives terms of curvature describes a transfer of energy and
momentum beyond that normally existing in curved spaces. This
interaction leads to the entropy production term in this modified
gravity. The FLT is formulated by employing the Wald's entropy
relation. We remark that an entropy production term is produced in
this work but no such term in present in GR, Gauss-Bonnet \cite{32},
Lovelock \cite{32} and braneworld \cite{37} theories of gravity.

The validity of GSLT has also been investigated in this work. We
have found that the GSLT holds with the conditions namely,
attractive nature of gravity and temperature being positive. In
fact, it is natural to assume the relation $T_{t}=bT_h$ and
proportionality constant $b$ can be considered as unity, implying
that the system is in thermal-equilibrium. Generally, the horizon
temperature cannot match the temperature of all energy sources
within the horizon, and the two mechanisms must experience
interaction for some interval of time ahead of achieving the
thermal-equilibrium. Moreover, the gravitational curvature-matter
coupling in $f(R,T)$ theory may produce the unscripted flow of
energy between the horizon and fluid. Also, the energy fluid of dark
components does not permit the effective gravitational constant to
be an approximate constant. In the limiting choice of
thermal-equilibrium, we assume that $T_t$ is very close to $T_h$. We
find that the GSLT is fulfilled in both phantom and quintessence
regimes of the universe which seems to be consistent with
refs.\cite{63}. Furthermore, we have also developed constraints on
some concrete $f(R,T)$ models corresponding to power law solution.
It is significant to remark that the equilibrium treatment of
thermodynamics in $f(R,T)$ theory would benefit from further study.

\vspace{0.25cm}

{\bf Acknowledgment}

\vspace{0.25cm}

The authors would like to thank the Higher Education Commission,
Islamabad, Pakistan for its financial support through the
\emph{Indigenous Ph.D. 5000 Fellowship Program Batch-VII}.

\end{document}